\documentclass{PoS}

\title{Heavy resonances and the electroweak effective theory}

\ShortTitle{Heavy resonances and the electroweak effective theory}

\author{\speaker{Ignasi Rosell}\thanks{We wish to thank the organizers for the pleasant conference. This work has been supported in part by the Spanish Government and ERDF funds from the European Commission (FPA2014-53631-C2-1-P, FPA2016-75654-C2-1-P, FPA2017-84445-P); by the Spanish Centro de Excelencia Severo Ochoa Program (SEV-2014-0398); by the Generalitat Valenciana (PROMETEO/2017/053); by the Universidad Cardenal Herrera-CEU (INDI16/10 and INDI17/11); and by the STSM Grant from COST Action CA16108. C.K. acknowledges the support of the Alexander von Humboldt Foundation. This manuscript has been authored by Fermi Research Alliance, LLC under Contract No. DE-AC02-07CH11359 with the U.S. Department of Energy, Office of Science, Office of High Energy Physics. Preprint numbers: IFIC/18-43, FTUV/18-1126, FERMILAB-CONF-18-652-T.}\\
        Departamento de Matem\'aticas, F\'\i sica y Ciencias Tecnol\' ogicas, Universidad Cardenal Herrera-CEU, CEU Universities, 46115 Alfara del Patriarca, Val\`encia, Spain \\
        E-mail: \email{rosell@uchceu.es}}
        
\author{Claudius Krause \\
Theoretical Physics Department, Fermi National Accelerator Laboratory, Batavia, IL, 60510, USA \\   E-mail: \email{ckrause@fnal.gov}}        
        
\author{Antonio Pich and Joaqu\'\i n Santos \\
IFIC, Universitat de Val\`encia -- CSIC, Apt. Correus 22085, 46071 Val\`encia, Spain\\   E-mail: \email{pich@ific.uv.es, joaquin.santos@ific.uv.es}}
    
    \author{Juan Jos\'e Sanz-Cillero \\ Departamento de F\'\i sica Te\'orica and UPARCOS,  Universidad Complutense de Madrid, E-28040 Madrid, Spain \\ E-mail: \email{jjsanzcillero@ucm.es}}

\abstract{Taking into account the negative results of direct searches for beyond the Standard Model fields and the consequent mass gap between Standard Model and possible unknown states, the use of electroweak effective theories is justified. Whereas at low energies we consider a non-linear realization of the electroweak symmetry breaking with a singlet Higgs and a strongly-coupled ultraviolet completion, at higher energies the known particles are assumed to be coupled to heavy states: bosonic fields with $J^P=0^\pm$ and $J^P=1^\pm$ (in electroweak triplets or singlets and in QCD octets or singlets) and fermionic states with $J=\frac{1}{2}$ (in electroweak doublets and in QCD triplets or singlets). By integrating out these heavy resonances, the pattern of next-to-leading order low-energy constants among the light fields can be studied. A phenomenological study trying to estimate the scale of these resonances is also shown.}

\FullConference{The 39th International Conference on High Energy Physics (ICHEP2018)\\
		4-11 July, 2018\\
		Seoul, Korea}

\begin{document}

\section{Introduction}

The LHC has confirmed that the Standard Model (SM) describes the electroweak (EW) scale very well and the existence of a Higgs-like particle, which couples following the SM predictions. Until now the experiments have not found evidences of new physics (NP): this mass gap between the EW and possible NP scales allows us to use effective field theories (EFTs) in order to analyze systematically the imprints of possible high-energy fields at low energies.

In EFTs the information about high energies is encoded in the low-energy constants (LECs). However, the information about low energies is described by the local operators. If the LECs are free parameters, the framework is model-independent but for some well-motivated assumptions (particle content, symmetries and power counting). The Higgs can be incorporated in two different ways: either by assuming that it forms an infrared doublet structure with the three Goldstone bosons of the EW symmetry breaking (EWSB) or without assuming any specific relation between them. The first one corresponds to the linear realization of the EWSB, usually called SM effective field theory (SMEFT). The second one, called EW effective theory (EWET), EW chiral Lagrangian (EWChL) or Higgs effective theory (HEFT), is a more general (non-linear) realization of the EWSB, where strongly-coupled scenarios are usually considered, and it is the one we follow in this work. 

At high energies the resonances can be incorporated by using a phenomenological Lagrangian which respects the EWSB pattern implemented in the SM, being this the main assumption of our approach. Integrating out these NP states, one recovers the EWET Lagrangian with its LECs given in terms of resonance parameters. Then, by fitting the LECs to experimental data, one can get the {\it imprints} of the fundamental EW Theory, the main aim of this kind of analysis. In Refs.~\cite{Pich:2015kwa,Pich:2016lew} we analyzed the {\it imprints} coming from colorless bosonic resonances and in Ref.~\cite{Krause:2018cwe} we enlarge the analysis to colored resonances, both of bosonic and fermionic kind.

\section{Low energies: the electroweak effective theory}

As pointed out previously, at low energies we use the EWET Lagrangian. The construction of the Lagrangian can be found in Ref.~\cite{Krause:2018cwe}, but we try to summarize them here:
\begin{enumerate}
\item Particle content: we consider the particle content of the SM. The Higgs is included as a scalar singlet with $m_h=125\,$GeV and for simplicity only a generation of fermions is included.
\item Symmetries: we assume the EWSB pattern, {\it i.e.}, $\mathcal{G}\equiv SU(2)_L\otimes SU(2)_R \rightarrow \mathcal{H}\equiv SU(2)_{L+R}$. The remaining symmetry $\mathcal{H}$ is called ``custodial'' symmetry, because it protects the ratio of the $W$ and $Z$ masses from receiving large corrections.
\item Power counting: we consider a low-energy expansion in powers of generalized momenta: $\mathcal{L}_{\mathrm{EWET}} = \mathcal{L}_{\mathrm{EWET}}^{(2)}+\mathcal{L}_{\mathrm{EWET}}^{(4)} +\dots\,$. Note that the operators are not ordered following their canonical dimensions (as it occurs in the SMEFT), but according to their chiral dimension, which reflects their infrared behavior at low momenta. There are two main differences comparing to previous works \cite{Buchalla:2013eza,Longhitano:1980iz}. Firstly, and assuming that the SM fermion are weakly coupled to the strong beyond-the-SM sector, we assign an $\mathcal{O}(p^2)$ to fermion bilinears. Secondly, assuming that no strong breaking of the custodial symmetry occurs, we assign an $\mathcal{O}(p)$ to the explicit breaking of this symmetry. Both suppressions are consistent with the phenomenology.
 \end{enumerate}
 

The next-to-leading order (NLO) Lagrangian can be organized by taking into account the parity ($P$) of the operators and the number of fermion bilinears~\cite{Krause:2018cwe}:
\begin{equation}
  \label{eq:L-NLO}
\mathcal{L}_{\mathrm{EWET}}^{(4)}  =
\sum_{i=1}^{12} \mathcal{F}_i\; \mathcal{O}_i  + \sum_{i=1}^{3}\widetilde{\mathcal{F}}_i\; \widetilde{\mathcal{O}}_i  
 +  \sum_{i=1}^{  8  } \mathcal{F}_i^{\psi^2}\; \mathcal{O}_i^{\psi^2}  + \sum_{i=1}^{  3   } \widetilde{\mathcal{F}}_i^{\psi^2}\; \widetilde{\mathcal{O}}_i^{\psi^2}
 + \sum_{i=1}^{10}\mathcal{F}_i^{\psi^4}\; \mathcal{O}_i^{\psi^4}  + \sum_{i=1}^{2}\widetilde{\mathcal{F}}_i^{\psi^4}\; \widetilde{\mathcal{O}}_i^{\psi^4}  ,
\end{equation}
where operators without (with) tilde refer to $P$-even ($P$-odd) ones and where we split the Lagrangian into bosonic, two-fermion and four-fermions operators, respectively.

\section{High energies: the resonance effective theory}

At high energies we consider the particle content of the SM plus bosonic fields with $J^P=0^\pm$ and $J^P=1^\pm$ (in electroweak triplets or singlets and in QCD octets or singlets) and fermionic states with $J=\frac{1}{2}$ (in electroweak doublets and in QCD triplets or singlets). Although the power counting explained in the previous section is not directly applicable here, a consistent organization can be followed by taking into account the integration of the resonances (in general $R\propto1/M_R^2$ and $\Psi\propto1/M_\Psi$ for the classical solution of the bosonic and fermionic resonances) and the additional suppression in case of fermionic resonances due to their weak coupling: only operators up to one resonance field and constructed with chiral operators of $\mathcal{O}(p^2)$ or lower are required. 

Once the resonance Lagrangian is set, the heavy resonances can be integrated out and the result can be organized in powers of momenta over the resonance masses. This procedure is standard in the context of EFT and allows us to determine the LECs of the NLO EWET Lagrangian in terms of resonance parameters, {\it i.e.}, $\mathcal{F}_i,\,\widetilde{\mathcal{F}}_i,\,\mathcal{F}_i^{\psi^2}, \widetilde{\mathcal{F}}_i^{\psi^2},\,\mathcal{F}_i^{\psi^4}$ and $\widetilde{\mathcal{F}}_i^{\psi^4}$ in (\ref{eq:L-NLO}) in terms of resonance couplings and masses~\cite{Pich:2015kwa,Pich:2016lew,Krause:2018cwe}.

As a first approach we studied in Ref.~\cite{Pich:2015kwa} the contributions to the LECs of $P$-even purely bosonic $\mathcal{O}(p^4)$ operators from $P$-even bosonic colorless resonance exchanges (without considering an explicit breaking of the custodial symmetry), once some short-distance constraints were taken into consideration, what allowed us to get predictions in terms of only a few resonance parameters. In Ref.~\cite{Pich:2016lew} we expanded our analyses to include bosonic and fermionic $\mathcal{O}(p^4)$ LECs from bosonic colorless resonance exchanges. Finally, in Ref.~\cite{Krause:2018cwe} we consider gluons, color octet fermion bilinears and fermionic resonances. In the future we plan to include more than one generation of quarks and/or leptons.

\section{Phenomenology}

In this section we try to glimpse the scale of the heavy states. The effects of vector and axial-vector resonances on the $S$ and $T$ parameters~\cite{Peskin:1990zt} were studied in Ref.~\cite{Pich:2012jv}, including NLO corrections. Current experimental bounds on these parameters and the requirement of a good high-energy behavior of the high-energy theory push the masses of the vector and axial-vector resonances to the TeV range~\cite{Pich:2015kwa,Pich:2012jv}. Note that possible scalar, pseudoscalar and fermion resonance contributions have not been considered yet. 

New interaction vertices as the ones of (\ref{eq:L-NLO}) have been analyzed in the literature: standard dijet and dilepton studies at LHC and LEP have searched for four-fermion operators containing light leptons and/or quarks~\cite{LHC_LEP}. In these analyses, the four-fermion LECs are usually expressed in terms of a suppression scale $\Lambda$ defined through $|\mathcal{F}^{\psi^4}_j| = 2\pi/\Lambda^2$. Currently, the most stringent (95\% CL) lower limits on this scale are:%
\begin{enumerate}
\item From dijet production~\cite{LHC_LEP}: $\Lambda\geq 21.8$~TeV from ATLAS, $\Lambda\geq 18.6$~TeV from CMS and $\Lambda\geq 16.2$~TeV from LEP.
\item From dilepton production~\cite{LHC_LEP}: $\Lambda\geq 26.3$~TeV from ATLAS, $\Lambda\geq 19.0$~TeV from CMS and $\Lambda\geq 24.6$~TeV from LEP.
\end{enumerate}
The previous bounds refer to four-fermion operators with the first and second generation of quarks. Some studies on four-fermion operators including top and bottom quarks can also be found~\cite{top_bottom}:
\begin{enumerate}
\item From high-energy collider studies~\cite{top_bottom}: $\Lambda\geq 1.5$~TeV from multi-top production at LHC and Tevatron, $\Lambda\geq 2.3$~TeV from $t$ and $t\bar{t}$ production at LHC and Tevatron and $\Lambda\geq 4.7$~TeV from dilepton production at LHC.
\item From low-energy studies~\cite{top_bottom}: $\Lambda\geq 14.5$~TeV from $B_s-\overline{B}_s$ mixing and  $\Lambda\geq 3.3$~TeV from semileptonic B decays.
\end{enumerate}

\end{document}